# Coherent Perfect Absorption in Chaotic Optical Microresonators for Efficient Modal Control

Xuefeng Jiang[1,†], Shixiong Yin[1,2,†], Huanan Li[1,3], Jiamin Quan[1], Michele Cotrufo[1], Julius Kullig[4], Jan Wiersig[4], and Andrea Alù[1,2,5*]

[1]Photonics Initiative, Advanced Science Research Center, The City University of New York, 85 St. Nicholas Terrace, New York, 10031, USA

[2]Department of Electrical Engineering, City College, The City University of New York, 160 Convent Avenue, New York, 10031, USA

[3]MOE Key Laboratory of Weak-Light Nonlinear Photonics, School of Physics, Nankai University, Tianjin 300071, China

[4]Institut für Physik, Otto-von-Guericke-Universität Magdeburg, Postfach 4120, D-39016 Magdeburg, Germany

[5]Physics Program, The Graduate Center, The City University of New York, 365 Fifth Avenue, New York, 10016, USA

[†] These authors contributed equally to this work.
* Correspondence to: aalu@gc.cuny.edu

## Abstract

Non-Hermitian wave engineering has attracted a surge of interest in photonics in recent years. One of the prominent phenomena is coherent perfect absorption (CPA), in which the annihilation of electromagnetic scattering occurs by destructive interference of multiple incident waves. This concept has been implemented in various platforms to demonstrate real-time control of absorption, scattering and radiation by varying the relative phase of the excitation signals. However, so far these studies have been limited to simple photonic systems involving single or few modes at well-defined resonant frequencies. Realizing CPA in more complex photonic systems is challenging because it typically requires engineering the interplay of a large number of resonances featuring large spatial complexity within a narrow frequency range. Here, we extend the paradigm of coherent control of light to a complex photonic system involving more than 1,000 optical modes in a chaotic microresonator. We efficiently model the optical fields within a quasi-normal mode (QNM) expansion, and experimentally demonstrate chaotic CPA states, as well as their non-Hermitian degeneracies, which we leverage to efficiently control the cavity excitation through the input phases of multiple excitation channels. Our results shed light on the universality of non-Hermitian physics beyond simple resonant systems, paving the way for new opportunities in the science and technology of complex nanophotonic systems by chaotic wave interference.

## Main

Non-Hermiticity and eigenvalue degeneracies in optical systems have been leveraged to achieve extreme wave phenomena[1,2], including parity-time symmetry[3,4], exceptional points[5], and CPA — the time reversal of coherent emission of radiation at the lasing threshold[6-8]. All these phenomena have been limited so far to simple systems involving single or few modes at separate and well-defined resonant frequencies[6-15]. In principle, they can emerge in any non-Hermitian system, provided that the coherent excitations are properly tailored to target the scattering matrix singularities for all input channels. In real-world settings, however, the challenge in realizing these features dramatically increases as more non-orthogonal resonances with complex spatial distributions arise in the real-frequency spectrum. In photonics, complex systems typically support numerous optical modes overlapping in their spectra and spatial distributions[16], making them hard to model interferometrically and by traditional reductionist approaches. Therefore, wave engineering in complex systems is conventionally realized by wave-shaping techniques[17-19], which is straightforward yet often requires significant time and energy for channel estimation and reconfiguration of wave modulators. Several attempts have been developed to address this limitation by coherent control of waves: CPA has been demonstrated in a medium with randomly-placed scatters[9], and more recently in a self-imaging Fabry-Perot cavity supporting arbitrary wavefronts[14]. Understanding and developing an efficient and general mechanism to exploit CPA and other exotic wave phenomena in the context of complex systems with highly spatial complexity may unleash exciting photonic opportunities in real-world scenarios, such as energy harnessing in chaotic system[20] and routing inside scattering media[21], chaotic laser imaging[22], chaotic optical communications[23,24], *etc*.

A prominent example of highly spatial complexity systems in nanophotonics is provided by chaotic optical microresonators, in which light exhibits chaotic ray dynamics such that a small deviation in the initial condition leads to an exponential divergence of trajectories[25,26]. The chaotic behavior of these complex photonic systems makes them a promising platform for fundamental physics studies, as well as for various photonic applications[27-32]. The characterization and control of the optical fields in such chaotic resonators are usually intricate[25]. Therefore, their response is typically modeled as a collective off-resonance background or a single low-quality-factor mode[33], even though in reality this response is associated with a very large number of individual resonances. Unlike conventional optical microresonators, however, the large modal density in chaotic photonic systems, as well as the complex overlap among different chaotic modes, makes it difficult to control and explore their overall dynamics, ruling out the application of new photonic applications limited to simple systems.

Here, we demonstrate that CPA, coherent control of light and non-Hermitian modal degeneracies can be extended to photonic systems with highly spatial complexity involving thousands of chaotic modes whose interplay can be tailored efficiently through multiple excitations. Without sophisticated cavity designs nor loss channel engineering, we show that chaotic CPA states arise massively in a stadium-shaped microcavity. Experimentally, by tuning the relative phase of a pair of coherent excitation beams, we demonstrate the emergence of numerous chaotic CPA states, as well as their degeneracy, within the telecommunication band. Leveraging them, we demonstrate the control of the far-field emission pattern emerging from the microresonator, opening a new direction for coherent control of chaotic complex systems.

We consider a silicon stadium microcavity buried in silica, formed by a rectangle of length $a$ – much larger than the wavelength – sandwiched between two semi-circles of radius $R = 2a/3$, as shown in Fig. 1a. The ray dynamics inside such a cavity is chaotic, featuring ergodicity without any stable periodic orbit[25]. Waveguide ports $1,2,3,...,N_{in}$ are assumed to carry coherent monochromatic signals exciting the stadium cavity. To demonstrate and describe the control over the excited fields via coherent excitations, we formally write the scattering matrix of the system as

$$\mathbf{S}(\omega) = \begin{pmatrix} \mathbf{S}_{in} & \mathbf{T}_{ri}^T \\ \mathbf{T}_{ri} & \mathbf{S}_{rad} \end{pmatrix}. \qquad (1)$$

The sub-matrix $\mathbf{S}_{in}$ describes the scattering within $N_{in}$ well-defined excitation channels, corresponding to the feeding waveguides, while $\mathbf{T}_{ri}$ couples the input waves towards the radiation continuum as mediated by the cavity. In general, the submatrices $\mathbf{T}_{ri}$ and $\mathbf{S}_{rad}$ associated with the radiation channels have infinite dimensions, spanning a continuous function space with outgoing-wave boundary conditions[34]. In most cases of physical interest, including our experiments, incoming waves from these radiation channels are absent, and $\mathbf{S}_{in}$ is sufficient to depict the control of the cavity excitation via coherent inputs (Supplementary Information). Even in the absence of material loss, $\mathbf{S}_{in}$ is generally non-Hermitian, since it describes a non-closed system, with energy spilling into radiation modes. We expand this sub-matrix into a set of QNMs $\psi_n$ supported by the chaotic resonator (as stacked on top of Fig. 1a), yielding the truncated scattering matrix (following an $e^{-i\omega t}$ notation)[35]

$$\mathbf{S}_{in} \approx -I - D(i\omega - i\Omega)^{-1}K^T, \tag{2}$$

where $K$ and $D$ relate the QNMs to the waveguide inputs $s_+$ and the corresponding reflection $s_-$, respectively, and the diagonal matrix $\Omega$ contains the QNM complex eigenfrequencies $\omega_n$, as labeled by crosses in Fig. 1b for the specific excitation geometry discussed in the following. A large number of QNMs are found around the normalized frequency range $Re(kR) = 25\sim29$ ($k$ is the wavenumber in free space), with quality factors varying from 100 to 1000.

As a prominent generalization of CPA in this scenario, the reflectionless scattering condition $s_- = \mathbf{S}_{in}s_+ = \mathbf{0}$ can be achieved upon a tailored coherent input $s_+$ at appropriate frequencies when $\text{eig}(\mathbf{S}_{in}) = 0$ (Ref. 36). For instance, in the case of excitation through two single-mode waveguides ($N_{in} = 2$), the eigenvalues of $\mathbf{S}_{in}$ can be generally expanded in QNMs in the form (Supplementary Information)

$$\sigma_{1,2}(\omega) \approx -1 - \frac{1}{2}\sum_n \frac{f_{11,n}+f_{22,n}}{i(\omega-\omega_n)} \pm \sqrt{\left[\frac{1}{2}\sum_n \frac{f_{11,n}-f_{22,n}}{i(\omega-\omega_n)}\right]^2 + \left[\sum_n \frac{f_{21,n}}{i(\omega-\omega_n)}\right]^2}, \tag{3}$$

where $f_{pq,n}$ ($p,q = 1,2$) are the expansion coefficients of $\mathbf{S}_{in,pq}$ given by Eq. (2). Given the chaotic nature of the cavity, we expect a large number of QNMs, and their nontrivial synergy captured by Eq. (3) enriches the possibility of control and manipulation of the scattering states through the two ports. Even under the assumption of symmetric excitation, when the two waveguides are connected symmetrically to the cavity, the spectrum of eigenvalues still exhibits rich physics: such symmetry requires $f_{11,n} = f_{22,n}$, hence $\sigma_\pm(\omega) \approx -1 + i\sum_n f_{n,\pm}/(\omega - \omega_{n,\pm})$ and $f_{n,\pm} = f_{11,n}^\pm \pm f_{21,n}^\pm$, where $\pm$ denotes the even (odd) symmetry of QNMs (Supplementary Information). The summation terms retrieved from simulations are shown for this scenario in Fig. 1c, where the indices $n$ correspond to different colors. The superposition of over ~1000 QNMs generates an abundance of zeros in the spectra of eigenvalues of the truncated scattering matrix, suggesting the emergence of a dense set of CPA states, indicated by the dashed lines in Fig. 1d obeying different symmetry conditions. By tuning the relative phase of the two excitations within a narrow wavelength range, we can readily access the large set of chaotic CPA states associated to the richness of QNMs. Interestingly, each chaotic CPA state is associated to a remarkably different chaotic field distribution. In particular, chaotic CPAs with even symmetry can be excited by $s_+ = (1,1)^T$, i.e., $\phi = 0$, if the overlap of even QNMs leads to $\sigma_+(\omega_{\text{CPA}+}) = 0$, while odd chaotic CPAs are found with $\phi = \pi$ if $\sigma_-(\omega_{\text{CPA}-}) = 0$. Their degeneracy (CPA$\pm$) can also emerge when both eigenvalues vanish at the same frequency, with exciting opportunities for efficient radiation control (Supplementary Information). Although the CPA symmetry is limited to even and odd nature because of the spatially symmetric setup, the dense number of chaotic CPAs with vastly different nature indicates that chaotic CPA degeneracies are also generally expected for asymmetric excitations (Supplementary Information).

In our experiment, a silicon stadium microresonator with $a = 9$ μm and $R = 6$ μm on a silicon-on-insulator (SOI) wafer (Methods) is fed by two nanoscale waveguides on the opposite sides of the stadium. As schematically shown in Fig. 2a, a tunable probe light within the telecom (around 1,550 nm) wavelength band is split into two counter-propagating beams, which are then coupled into the microresonator through Ports 1 and 2. The reflection signals are collected from Ports 3 and 4 (Methods). The relative phase between the two probe laser beams is modulated and scanned by a fiber electro-optical modulator (EOM). To experimentally observe the excited fields, a silicon barrier with width of 500 nm encircling the cavity (50 um away from the boundary of the stadium) was fabricated on chip. The barrier partially scatters the chaotic emission fields towards a near-infrared (NIR) camera placed above the chip. A scanning electron microscope (SEM) image of a fabricated device, including both stadium cavity and silicon barrier, is shown in Fig. 2b. We fabricated two 5-$\mu$m-wide notches on the silicon barrier to allow the nanoscale waveguides to pass through. The scattering intensity at a particular polar angle on the silicon barrier measured by the NIR camera is proportional to the emission intensity at the same angle. An example of measured scattering imaging is shown in Fig. 2c, in which we observe two bright spots in the middle of the image due to the strong scattering of the input beams reflected at the waveguide-cavity junctions.

To observe the chaotic CPA states, we first measured the reflection spectra for different relative phases ($\Delta\phi$) of the two excitations, by applying different constant voltages on the EOM, as shown in Fig. 3a. Many reflection zeros can be found within the narrow wavelength range 1520-1545 nm, validating our prediction of a dense number of chaotic CPA states. For instance, around 1524 nm (red shaded area) the normalized reflection level decreases from its peak to zero as $\Delta\phi$ changes from 0 to $\pi$, indicating a CPA− state for which the reflection at the port vanishes under excitation with opposite relative phases. The opposite trend (CPA+) is observed around 1530 nm (blue shaded area). Remarkably, at some wavelengths, such as at 1542 nm, the reflection remains near zero for all relevant phases, indicating the emergence of a non-Hermitian degeneracy (CPA±) of both chaotic CPA states. We investigated the phase dependence of the reflection at the ports by scanning the EOM voltage for fixed probe laser wavelength, as shown in Fig. 3b,c. In the scenario of CPA−, the measured reflection (red squares fitted by the solid curve in panel B) follows a $\cos^2(\Delta\phi/2)$ variation, reaching zero at $\Delta\phi=\pi$, while the reflection for CPA+ (blue ones in panel C) follows $\sin^2(\Delta\phi/2)$ and reaches zero at $\Delta\phi=0$, consistent with coherent control of absorption in single-mode or few-mode CPA systems, but here extended to a chaotic scenario with a large number of chaotic CPA states (Supplementary Information).

To further confirm the coherent control of chaotic field excitation, we compared the retrieved wave properties to ray chaotic dynamics using the Poincaré surface of section (SOS) and the Husimi function. Poincaré SOS records the angular momentum $\sin\chi$ and polar angle $\theta$ at each reflection of a ray, which in our stadium microcavity exhibits discrete points without any periodicity, highlighting the chaotic nature of the ray dynamics. The wave properties are projected by overlapping the boundary fields onto a coherent state (a minimal-uncertainty wave packet), representing the quasi-probability distribution of a quantum state in phase space[25,37]. The calculated Husimi functions for CPA− and CPA+ are shown in Fig. 3d,g, with the Birkhoff coordinates defined in the inset of Fig. 3c. Here we open a leaky "window" at the center of each plot, denoted by a green solid line located at $\theta = \pi$ and spanning $|\sin\chi| < 0.4$, at which the total internal reflection condition is violated (Supplementary Information). The amplitude of the Husimi functions in this

window, as shaded in the juxtaposed line plots, indicates the emerging probabilities to the left waveguide. For $\Delta\phi=\pi$ for the CPA− state (Fig. 3e), the function intensity remains relatively small in the leaky window, while at $\Delta\phi=0$, as shown in Fig. 3d, it reaches a peak at $\sin\chi = 0$, consistent with the reflection extrema in Fig. 3b. In the CPA+ case, the amplitude of the Husimi functions (Fig. 3f,g) exhibits the opposite behavior, validating the measured reflection in Fig. 3c as well. These Husimi functions for different CPA states confirm that the interplay of different QNMs can be efficiently tailored and controlled by tuning the relative phase of the probe light, providing a direct tool to control the chaotic fields.

Such coherent control of light in the chaotic microresonator can be used to tailor in real-time the radiated fields from the cavity. We experimentally measured the radiation patterns by taking snapshots with an NIR camera while scanning the relative phase of the excitation fields. The scattering intensity at a particular polar angle can be retrieved from the corresponding pixel intensities recorded by the camera. Figure 4a,b present the intensity of scattered fields upon relative phase changes ($\Delta\phi$) and at different polar angles $\Theta$ for CPA− and CPA+, respectively. It is clear that, by operating at a chaotic CPA, it is possible to widely control the angular emission pattern through the coherent superposition of excitation fields. We also integrated the scattered fields by the silicon barrier for different relative phases ($\Delta\phi$), as shown in Fig. 4c,d, corresponding to CPA− and CPA+, respectively. As expected, the integrated scattered intensities as a function of relative phase change $\Delta\phi$ show opposite behaviors for CPA+ and CPA− compared to the reflection signals shown in Fig. 3b,c, which can be explained by energy conservation (Supplementary Information): at a minimum of reflection, we find maximum radiation leakage from the chaotic microresonator, confirming that the relative phase of excitations can tailor in real-time not only the angular pattern but also the level of emission from the cavity.

Finally, we focus on a degenerate CPA± state with excitation wavelength around 1542 nm. As shown in Fig. 5a, the reflection at this wavelength remains close to zero when varying the relative phase between the beams. Meanwhile, the integrated scattered intensities (Fig. 5b) remain constant as the relative phase changes. It is worth noting that the power level of the scattered intensities in the case of CPA± (Fig. 5b) is larger than the one of CPA− (Fig. 4c) and CPA+ (Fig. 4d) at every phase, which are normalized to the same value across the different panels, due to the uniform reflectionless feature of CPA±. Figure 5c shows the measured scattered fields upon different relative phases and at different azimuthal angles. At three different relative phases ($\Delta\phi = 0$, $\pi/2$, and $\pi$), the scattering patterns exhibit noticeable differences (Fig. 5d), attributed to different QNMs contributions and induced chaotic fields inside the cavity, demonstrating efficient emission control through wave interference in our chaotic system (see Supplementary Information for simulations). Non-Hermitian degeneracies of chaotic CPAs form an ideal platform to obtain efficient control of emission with no reflection[38] in nanophotonic structures.

Our results demonstrate that it is possible to extend the concept of CPA and non-Hermitian degeneracies to complex optical systems by accurately controlling the excitation of chaotic modes, paving the way towards new opportunities in vital photonic applications based on chaotic systems[25], including quantum chaos, nonlinear optics, lasers, and cavity optomechanics. We have also demonstrated that, due to the large density of chaotic CPA states, it is possible to induce degenerate chaotic CPAs, for which the reflection at the port remains minimal, but the modal distribution is widely modified as we scan the phase difference between excitation fields (see Supplementary Information). This operational principle provides an ideal platform to investigate broadband energy storage and anti-reflection phenomena in real-world scenarios involving complex structures. We also envision the extension of these concepts to asymmetric excitations, and multiple (>2) excitation schemes, further enriching the degree of chaotic CPAs and real-time control over the induced chaotic fields (Supplementary Information). Similar concepts can be extended to other chaotic systems, such as chaotic microlasers[39], cavity optomechanical systems[40,41], and chaos-based optical communications[23,24]. These mechanisms may also be explored in other complex systems, including not only random scattering media[9,38], and random lasing[42,43], but also communication systems, and electronic systems.

## Methods

**Device Fabrication Procedure**

The stadium microresonators were fabricated using standard nano-fabrication techniques on silicon-on-insulator (SOI) wafers, with the thicknesses of the device layer and buried oxide layer of around 220 nm and 3000 nm, respectively. Hydrogen silsesquioxane (HSQ) was used as a negative tone inorganic electron beam resist for patterning the microresonator, waveguide, and the barrier structures. HSQ powders (Applied Quantum Materials Inc., H-SiOx) were dissolved in methyl isobutyl ketone (MIBK) to produce HSQ solution with a concentration of 10%, which was then spun coated on a cleaned SOI wafer to form a 200-nm-thick resist layer. The resist was then soft baked at 85 °C for two minutes, and the designed structures were then written by electron beam exposure at a beam current of 1 nA and voltage of 100 keV. A 195°C post exposure baking was then performed for 2 min to enhance the contrast properties of the film. The patterned wafer was then developed in 25% tetramethylammonium hydroxide (TMAH) for 70 seconds and rinsed in deionized water and isopropyl alcohol (IPA) for 60 seconds and 20 seconds, respectively. The exposed HSQ served as an etching mask for the subsequent inductively coupled plasma (ICP) etching process with gases of $C_4F_8$ and $SF_6$. A 2-μm-thick silica layer was then deposited, via plasma-enhanced chemical vapor deposition (PECVD), to form a cladding and protection layer. Edge couplers were fabricated by cutting the chip, at a direction perpendicular to the waveguide, with a dicing saw (Disco DAD3220). The cutting was done in correspondence of a specially designed tapered waveguide position with a width of around 100 nm, which possesses an optimized coupling efficiency to a Gaussian beam in the 1550 nm wavelength band and with a diameter of 5 μm[44]. The cut chip edges are then polished by a homemade chemical mechanical polishing machine to fine-tune the width of the edge tapered waveguide and to form a smooth end surface of the waveguide.

**Experimental Measurement**

In experiments, the silicon stadium microresonator is coupled to two nanoscale waveguides, with widths of around 200 nm, located at the opposite sides of the stadium as shown in Fig. 2b. Each one of these two waveguides (marked as Port 1 or 2 in Fig. 2a) couples evanescently to another waveguide (Port 3 or 4) to form a 50:50 coupler for the detection of reflected signals. As schematically shown in Fig. 2a, a tunable probe light (within the telecom wavelength band around 1,550 nm) is split into two counter-propagating beams by a fibre coupler, which are then coupled into two tapered edge-couplers (Ports 1 and 2) on both sides of the SOI wafer, by two individual lensed fibres (OZ Optics Ltd). The lensed fibres produce a focused spot with a diameter of around 5 microns. Meanwhile, the reflection signal from Port 3 or 4 is collected and detected by a multimode fibre and a photoreceiver, respectively. The intensities and polarizations of these two input beams are independently adjusted by a fibre attenuator and two fibre polarization controllers, respectively. The phases of the two probe laser beams are matched by a fibre variable delay line (Newport, F-VDL-2-6-FA-S). Then, the relative phase of these two beams is modulated by a fibre electro-optical phase modulator (Thorlabs, LN65S-FC), which can be scanned/tuned by applying a triangle wave through a function generator.

**Numerical Simulations**

All the full-wave simulations are performed with COMSOL Multiphysics. To guarantee the single-mode waveguide excitations and reflections, we exploit the spatial symmetry of our geometry, and set the symmetric axis as a perfect magnetic conductor. The eigenfrequencies in Fig. 1b are computed by the eigenfrequency solver with the presence of two waveguides of width $w = \pi/(nk_0)$ where $n = 3.48$ and $k_0 R = 25$.

## Data availability

The data that support the findings of this study are available from the corresponding authors upon reasonable request.

## Acknowledgements

We thank Zhan Li at Stevens Institute of Technology for polishing the edge couplers. This work was supported by the Air Force Office of Scientific Research and the Simons Foundation.

## Author contributions

X.J., S.Y., and A.A. conceived the project; X.J. fabricated the device and designed the experiments; X.J. and S.Y. performed the experiments and analyzed experimental data with help from J.Q., M.C., and A.A.; Theoretical studies and simulations were performed by S.Y. with help from X.J., H.L., J.K., J.W., and A.A; All authors discussed the results; A.A. supervised the project; X.J, S.Y., and A.A. wrote the manuscript with input from all co-authors.

**FIGURES**

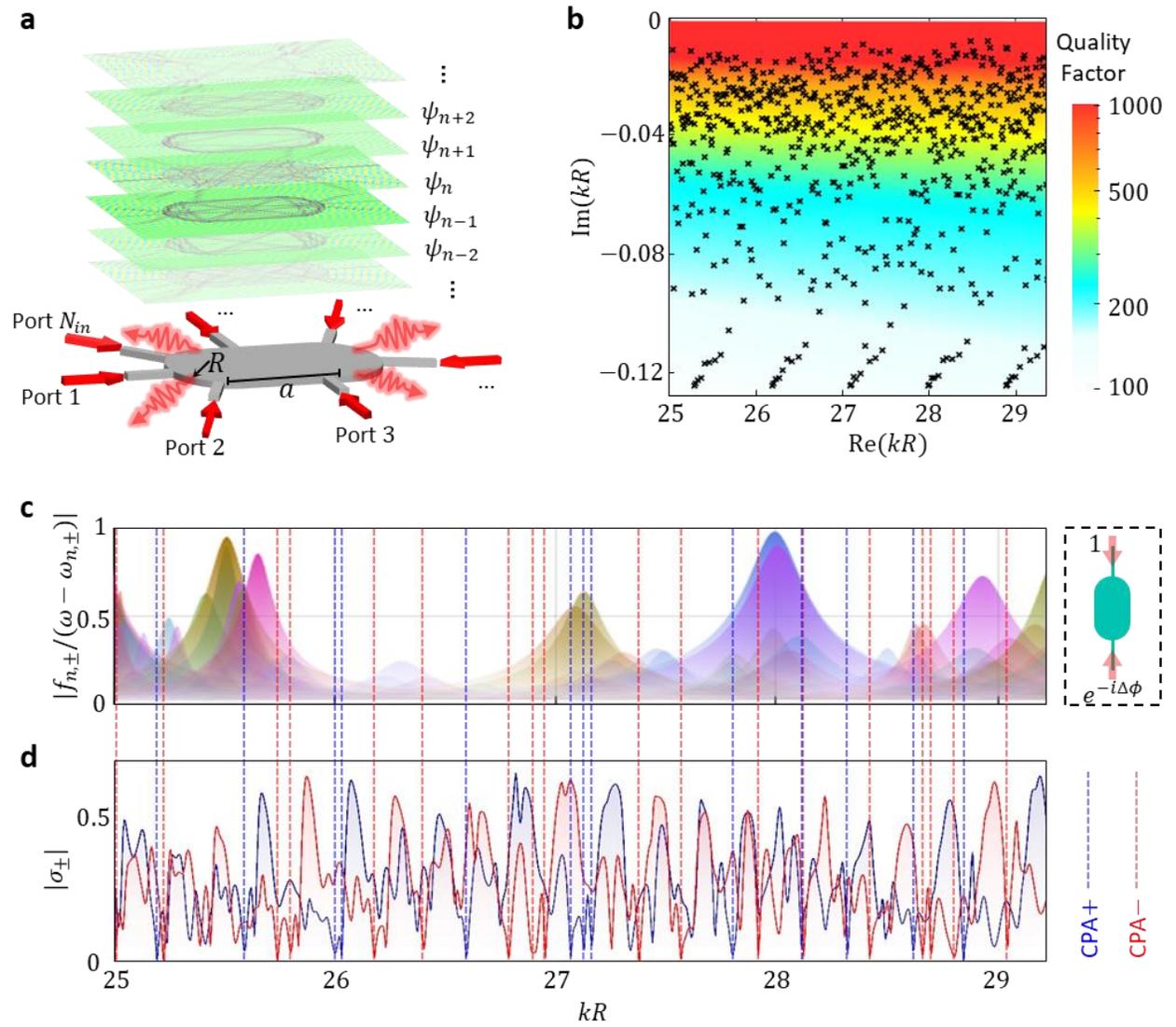

**Fig. 1. Modal analysis of coherent perfect absorption (CPA) in a chaotic optical microresonator.** (**a**) A silicon stadium billiard microcavity is excited by $N_{in}$ waveguide ports. The energy leakage is modeled with radiation channels (red wavelets). The supported quasi-normal modes (QNMs) are sketched on top. (**b**) Retrieved complex eigenfrequencies of the waveguide-fed stadium microcavity shown in panel **a**, normalized to the complex variable *kR*. The coloured background indicates the quality factor of the QNMs. (**c**) The contribution of each QNM (in different colours) to the eigenvalues of the truncated scattering matrix for the two symmetric waveguide ports, as shown in the subpanel on the right. (**d**) Spectra of two

eigenvalues of the truncated scattering matrix. Numerous zeros are observed contributed by a large number of QNMs, corresponding to chaotic CPAs with even (blue dashed lines) and odd (red dashed lines) symmetry.

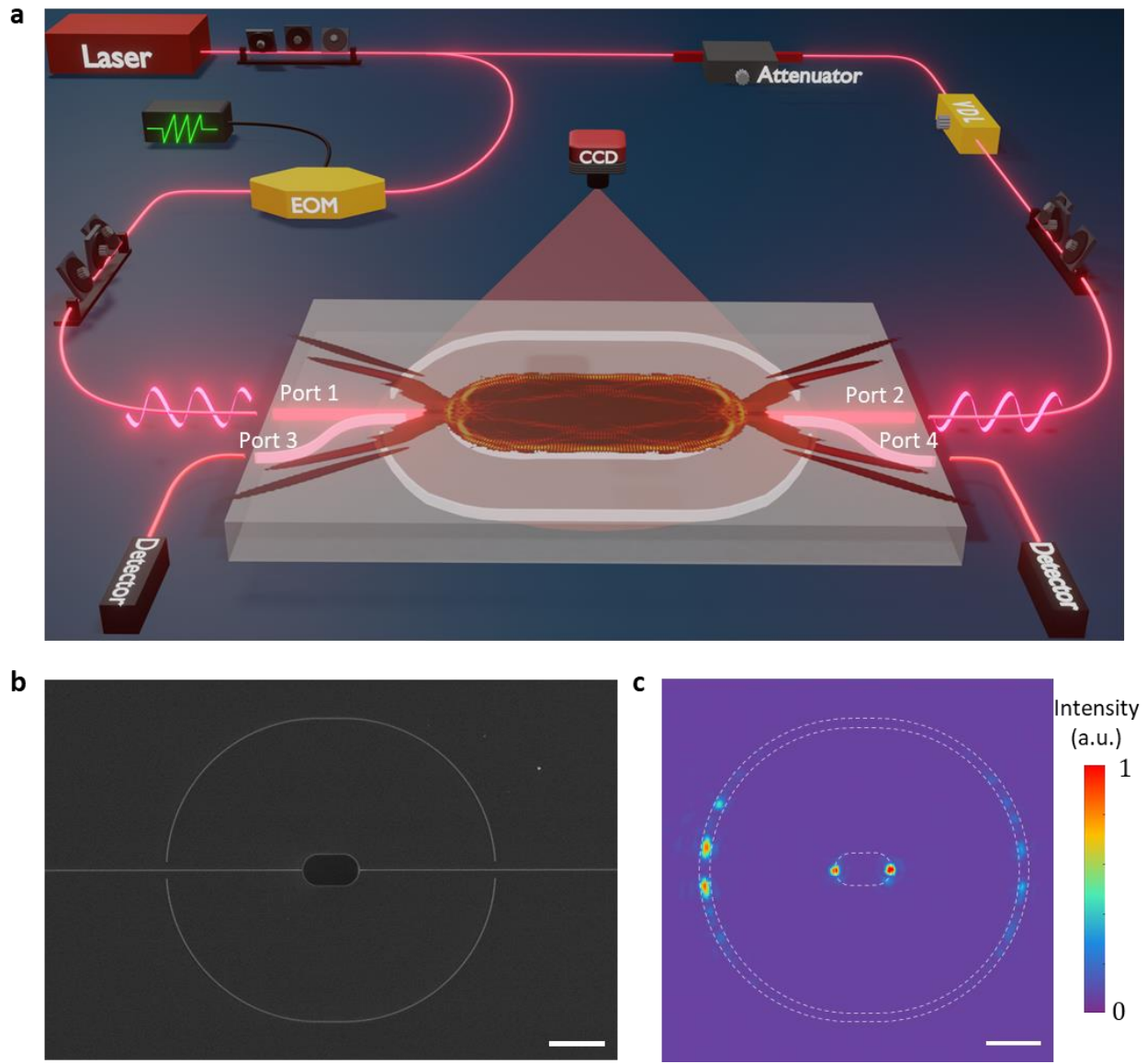

**Fig. 2. Experimental setup and chaotic optical microresonator system.** (**a**) Schematic diagram of the experimental setup. EOM: electro-optics modulator, VDL: variable delay line. (**b**) Scanning electron microscope (SEM) imaging of a fabricated device. (**c**) Near-infrared (NIR) camera imaging of the scattering of the silicon barrier. White dashed curves denote the boundaries of the stadium cavity, the silicon barrier as well as the nano-waveguides shown in panel **b**.

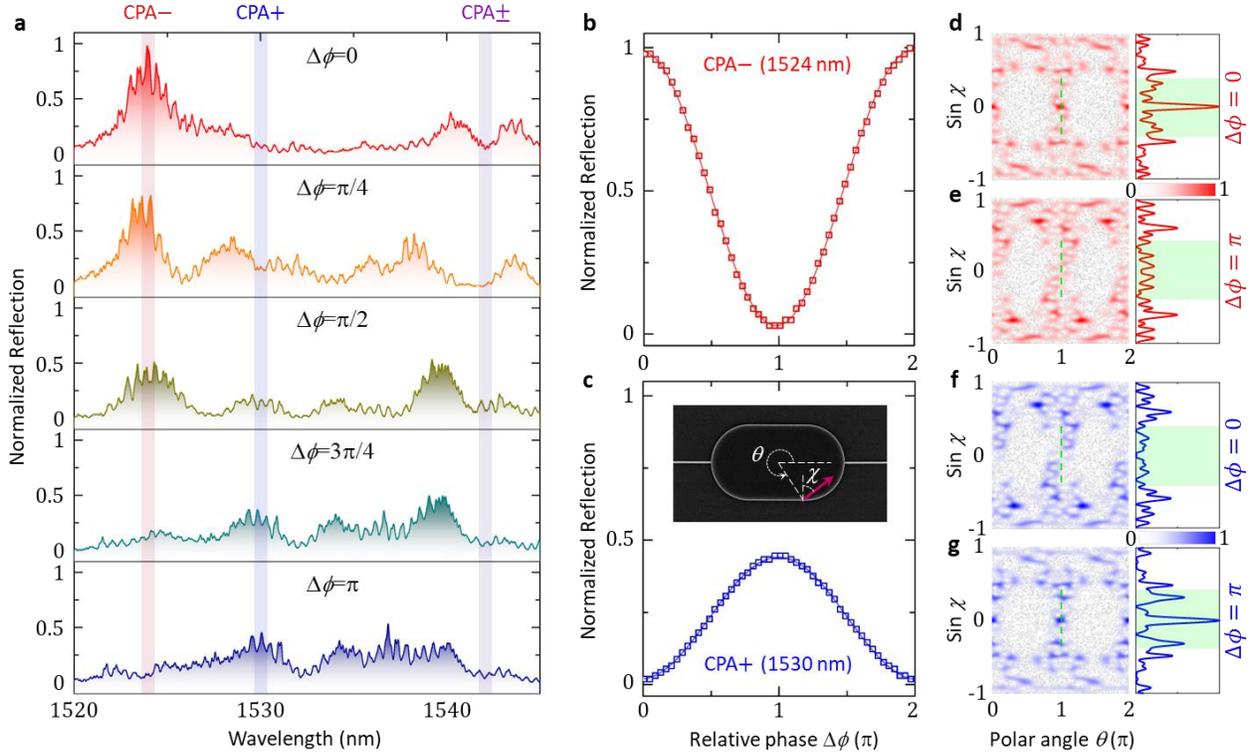

**Fig. 3. Experimental results for chaotic CPA.** (**a**) Reflection spectra for different relative phases $\Delta\phi$ of the exciting fields. (**b** and **c**) Phase modulation of reflection of a CPA− state at 1524 nm (**b**) and a CPA+ state at 1530 nm (**c**). The inset in panel **c** shows an SEM of the device and illustrates the definition of the Birkhoff coordinates ($\theta$ and $\chi$) for the Husimi functions. (**d** and **e**) Simulated Husimi functions with Poincare surface of section (SOS) as a background at CPA− with relative phases $\Delta\phi=0$ (**d**) and $\Delta\phi=\pi$ (**e**). The insets show the Husimi intensity at $\theta = \pi$, where the left waveguide locates. (**f** and **g**) Simulated Husimi functions with SOS background at CPA+ with relative phases $\Delta\phi=0$ (**f**) and $\Delta\phi=\pi$ (**g**). The insets show the Husimi intensity at $\theta = \pi$. The green solid lines in the Husimi functions and green shadow areas in Husimi intensity at $\theta = \pi$ represent the emerging probabilities to the left nanoscale waveguide.

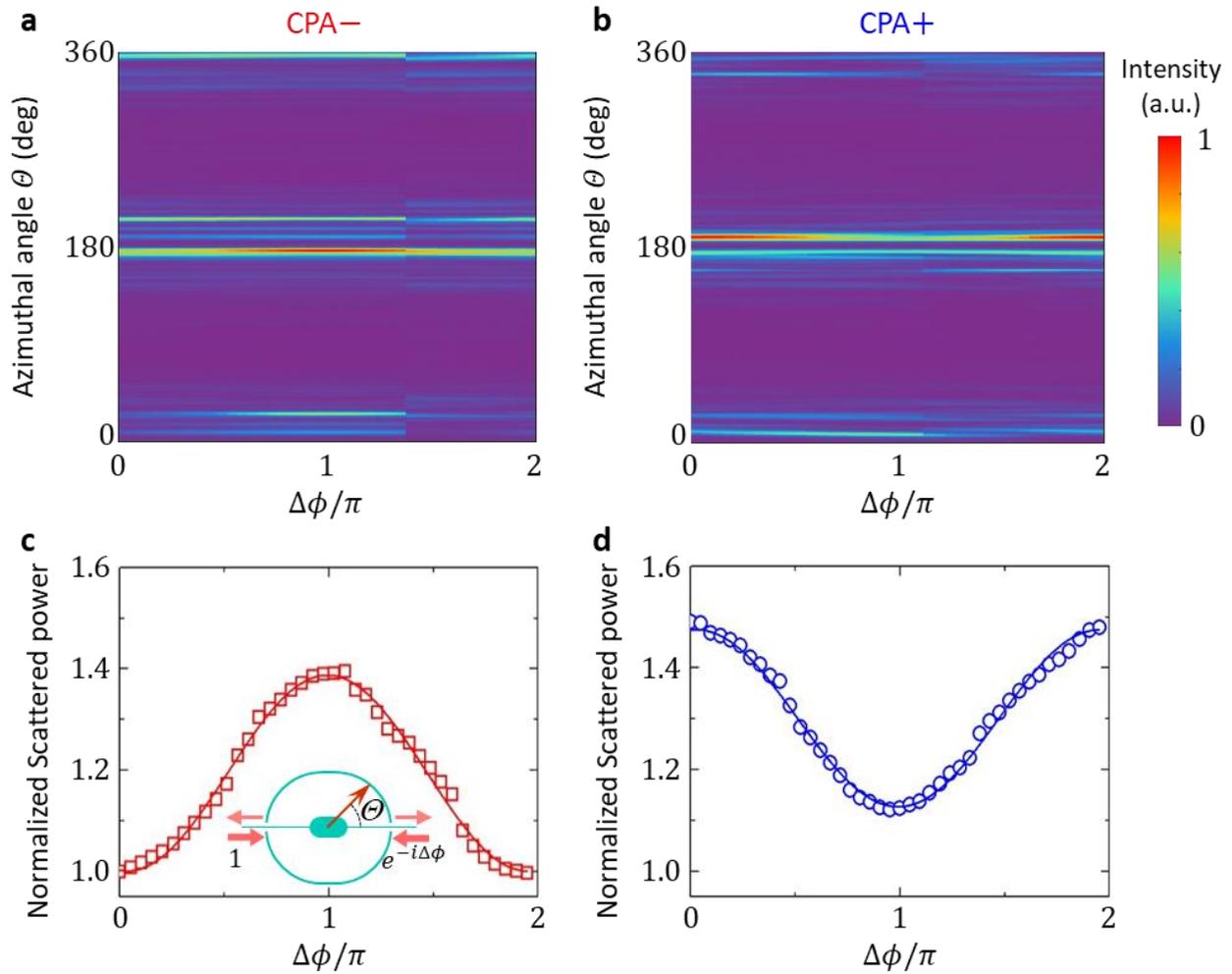

**Fig. 4. Experimental measurement of emission field intensities.** (**a** and **b**) Intensities of the scattering fields on the silicon barrier as a function of relative phase changes ($\Delta\phi$) and polar angles for CPA− (**a**) and CPA+ (**b**). (**c** and **d**) Integrated scattering fields on the silicon barrier at different relative phase changes ($\Delta\phi$), for CPA− (**c**) and CPA+ (**d**). The inset illustrates the definition of azimuthal angle $\Theta$ and relative phase $\phi$ in panels **a** and **b**.

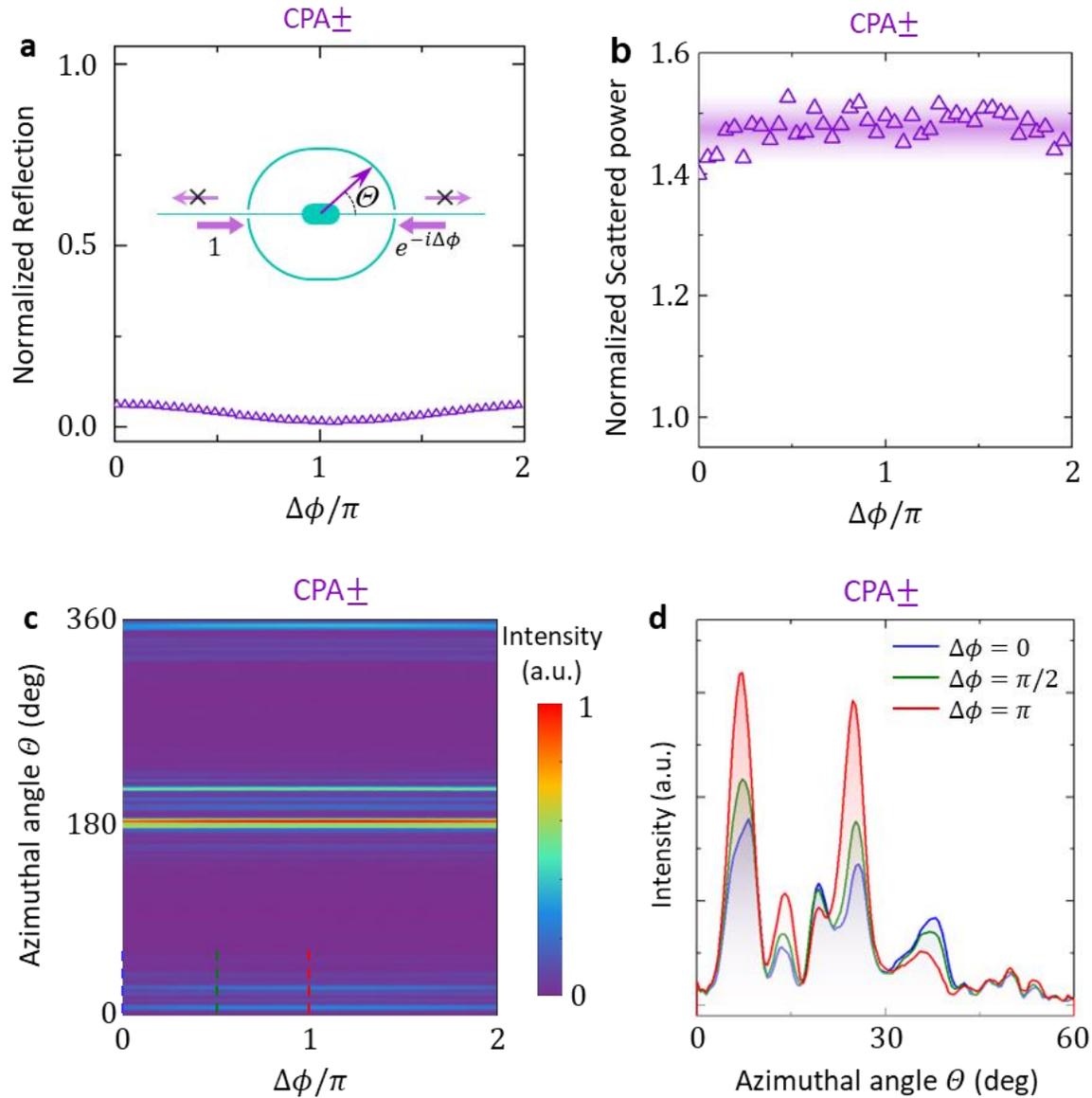

**Fig. 5. Experimental results of the degeneracy of chaotic CPAs.** (**a**) Reflection level of a CPA± state at 1542 nm versus the relative phase $\Delta\phi$ between the beams. The inset illustrates the definition of azimuthal angle $\Theta$ and relative phase $\Delta\phi$ used in panels **c** and **d**. (**b**) Integrated scattered power on the silicon barrier versus the relative phase. (**c**) Scattered power on the silicon barrier as a function of relative phase and polar angles for CPA±. The dashed lines mark the azimuthal angles and phases of data in panel **d**. (**d**) Scattered intensitiy distribution versus the azimuthal angle $\Theta$ for three different input phases ($\Delta\phi = 0$, $\pi/2$, and $\pi$).